# Theoretical investigation of real-fluid effects on the oxidation of hydrogen and syngas in high-pressure flow reactors

Mingrui Wang[a], Tianzhou Jiang[a], Ting Zhang[a], Hongjie Zhang[a], Hongqing Wu[a], Rongpei Jiang[b], Yang Li[c], Song Cheng[a,*]

[a] Department of Mechanical Engineering, The Hong Kong Polytechnic University, Kowloon, Hong Kong SAR, China
[b] Beijing Institute of Aerospace Testing Technology, Beijing, China
[c] National Key Laboratory of Solid Propulsion, School of Astronautics, Northwestern Polytechnical University, Xi'an, China

*Corresponding author.
Email: songryan.cheng@polyu.edu.hk
Song Cheng
Phone: +852 2766 6668

**Abstract:**

Investigating the high-pressure pyrolysis and oxidation of hydrogen and syngas is essential for developing accurate combustion models under real engine conditions, enabling safer, cleaner, and more efficient engine systems. Accordingly, flow reactors have been increasingly used in these fundamental combustion research to probe into fuel pyrolysis and oxidation chemistry at high-pressure conditions. This type of reactor has also been fundamental to the development and validation of chemistry models via chemical kinetic modeling. However, previous chemical kinetic modeling for pyrolysis and oxidation in high-pressure flow reactors has almost been completely conducted based on ideal gas assumption, although the ideal gas assumption might fail at high-pressure conditions when real-fluid behaviors become pronounced. Elucidating whether this is the case is urgent given the high demand for high-pressure combustion and propulsion technologies. Unfortunately, this has been seriously overlooked in the past. Although there have been a few attempts to quantify real-fluid effects in high-pressure flow reactors, the theoretical foundation used therein has been proven to be insufficient for combustion studies. Toward this, this study establishes a first-of-its-kind real-fluid modelling framework for high-pressure flow reactors, where the physical molecular interactions in real fluids are represented via coupling *ab initio* intermolecular potentials with high-order Virial equation of state and further with real-fluid thermochemistry, real-fluid chemical equilibrium, and real-fluid conservation laws of species, mass, and momentum. Using the developed framework, real-fluid effects in high-pressure flow reactors are quantified through case studies of hydrogen and syngas oxidation, covering pressure of 10-500 bar, temperature of 601-946 K, equivalence ratio of 0.0009-12.07, and dilution ratio of 5.9%-99.5%. The results reveal the strong real-fluid effects in high-pressure flow reactors, where ignoring them can lead to considerable errors in the simulated mole fractions (as high as 100%), which are higher than typical levels of measurements uncertainty in flow reactors. These errors can lead to misinterpretation of the fundamental oxidation chemistry in high-pressure flow reactors and pose significant errors in the developed chemistry models, which should be adequately accounted for in future high-pressure flow reactor studies (e.g., via the real-fluid modeling framework developed in this study).

**Keywords:** High-pressure flow reactor; real-fluid modeling framework; high-order Virial equation of state; *ab initio* intermolecular potentials; high-pressure hydrogen and syngas oxidation

**CRediT authorship contribution statement**

**Mingrui Wang**: Writing – original draft, Conceptualization, Methodology, Formal analysis, Investigation. **Tianzhou Jiang**: Formal analysis, Investigation. **Ting Zhang**: Formal analysis, Investigation. **Hongjie Zhang**: Formal analysis, Investigation. **Hongqing Wu**: Formal analysis, Investigation. **Rongpei Jiang**: Formal analysis, Investigation. **Yang Li**: Formal analysis, Investigation. **Song Cheng**: Writing – original draft, Conceptualization, Methodology, Formal analysis, Investigation, Writing – review & editing, Supervision, Funding acquisition.

# 1. Introduction

Flow reactors (FRs), an important type of facility with a pressure- and temperature-controlled reaction chamber, have been extensively used in isothermal chemical processes [1]. Due to their advantages, FRs are now widely applied for both fundamental research and industrial applications concerning continuous flow, e.g., flow microwave reactors [2], microchip photo reactors [3], and electrochemical flow cells [4]. In recent years, high-pressure FRs, especially those working under supercritical conditions, have attracted increasing attention. Supercritical fluids feature unique properties not obtainable in the ambient environment (e.g., relatively low viscosity, high diffusivity, and no surface tension) [5], and therefore exhibit superiority in some specific fields, such as synthesis [6], hydrogenation [7], and oil recovery [8] in FRs.

In fundamental combustion, tubular flow reactors are extensively used for studying pyrolysis and oxidation chemistry due to their operational flexibility, where temperature, pressure, and gas fluid residence time can be carefully controlled. In comparison with batch reactors, such as rapid compression machines and shock tubes, which are essentially transient reactors, flow reactors can be considered as steady-state reactors if controlled well. To this end, FRs aiming to study fundamental combustion chemistry typically run on extremely lean and diluted mixtures, so that the temperature across the reactor can be maintained constant, hence a steady state.

There have been several tubular flow reactors that have been specifically designed to study high-pressure pyrolysis and oxidation. In 1991, a variable-pressure turbulent flow reactor capable of operating temperatures approaching 1200 K and pressure of 0.2-20 atm was designed by Vermeersch [9] at Princeton University, and was applied to investigate gas-phase pyrolysis and oxidation of $C_2H_5OH$ [10, 11], DME [12, 13], surrogate fuel [14], etc. Later in 2008, a laminar flow reactor operating at higher pressures (up to 100 bar) and 450-900 K was developed at the Technical University of Denmark [15], and has been widely used to validate chemical kinetic models and investigating the high-pressure oxidation characteristics of syngas [16], $NH_3$ [17], $CH_4$ [18], $C_2H_6$ [19], $C_3H_8$ [20], DME [21], n-Heptane [22], etc. Subsequently, a plug flow reactor for investigating the oxidation chemistry of fuels at up to 50 bar and 1000 K was developed by Lu et al. [23] at The University of Melbourne. More recently, an ultra-high-pressure plug-flow reactor was designed by Li et al. [24] to investigate hydrogen oxidation in supercritical $H_2O/CO_2$ mixtures at 236 bar and 778-873 K.

In previous fundamental studies on pyrolysis and oxidation conducted with high-pressure flow reactors [9-24], the developed chemical kinetic model is refined to achieve better consistency between modelling and experimental results, which is usually achieved by adjusting rate constants and adding new reactions or intermediates of the chemical kinetic model. On the other hand, it is important to note that in dense phases, intermolecular interactions cause deviations in fluid properties from the ideal fluids, typically referred to as real-fluid effects at high pressures (e.g., shifted thermochemistry and chemical equilibrium with pressure increasing [25]). However, these real-fluid effects are mostly ignored during modelling pyrolysis and oxidation in high-pressure flow reactors [26], which might lead to errors in the modeling results, thus unreliable validation and optimization of chemical kinetic models. Toward this, in recent years, several real-fluid equations of state (EoSs), traditionally the cubic EoSs, have been used to account for real-fluid effects during high-pressure oxidation in high-pressure flow reactors. For instance, Li et al. [27] adopted Peng-Robinson (PR) EoS to calculate the mixture density when developing kinetic models for hydrogen oxidation in supercritical $H_2O/CO_2$ mixtures at 236-250 bar. Manikantachari et al. [28] developed real-fluid thermochemistry by incorporating Soave-Redlich-Kwong (SRK) EoS into CHMEKIN [29] when modelling methane oxidation at 300 atm in plug-flow reactors. While these real-fluid cubic EoSs require accurate predefined interaction coefficients to sufficiently determine the real-fluid properties [30], such real-fluid EoSs are empirical in essence. In fact, it has already been demonstrated that these cubic EoSs become inaccurate under the conditions where experimental data are unavailable, particularly when more species/radicals are involved (e.g., combustion processes). Although there have been some attempts on capturing real-fluid behaviors in high-pressure flow reactors, almost all modeling of high-pressure flow reactor experiments today are still conducted based on ideal gas assumption.

Recently, the high-order Virial EoS derived based on statistic mechanics [31], where the equation parameters (i.e., Virial coefficients) are computed directly from *ab initio* intermolecular potentials, was proven to outperform ideal gas EoS and cubic EoSs in predicting thermochemical properties at a wide pressure range, covering 1-1000 bar [32, 33]. The superiority of high-order Virial EoS was confirmed in predicting enthalpy, heat capacity, partial molar volume, and partial molar enthalpy. The real-fluid thermochemistry and the developed real-fluid chemical equilibrium are subsequently incorporated into the real-fluid combustion conservation laws to establish real-fluid modelling frameworks for high-pressure fundamental combustion in shock tubes [32, 33], rapid compression machines [34], and jet-stirred reactors [35, 36]. These studies found that ignoring real-fluid effects during the modeling of high-pressure fundamental combustion experiments can lead to an error of up to 65% in the modeling results at the experimental conditions investigated and lead to contradictory validation results for chemical kinetic models, making it necessary to quantify the real-fluid effects in high-pressure flow reactors.

With such awareness, this study aims to establish a real-fluid modelling framework for the pyrolysis and oxidation in high-pressure flow reactors that couples *ab initio* intermolecular potential, real-fluid thermochemistry, real-fluid chemical equilibrium, and real-fluid conservation laws of species, mass, and momentum. This new modelling framework is utilized to reveal and quantify the real-fluid effects in typical high-pressure flow reactors, demonstrated via case studies of hydrogen and syngas. Based on the quantified real-fluid effects, the errors introduced into modeling results without considering non-ideal behaviors are revealed and further quantified, thereby discussing how these errors can affect chemistry model validation against high-pressure flow reactor experiments. Finally, the contribution of real-fluid chemical equilibrium to the overall real-fluid effects in high-pressure flow reactors is investigated.

## 2. Methodologies

*2.1 Real-fluid modeling framework for high-pressure flow reactors*

In this study, the real-fluid modeling framework for high-pressure flow reactors is developed based on high-order Virial EoS coupling *ab initio* intermolecular potentials, following a similar logic as the real-fluid modeling frameworks that the authors' group has previously developed for high-pressure rapid compression machines [34], shock tubes [32, 33] and jet-stirred reactors [35, 36]. The details of the implementation of *ab initio* intermolecular potentials into high-order Virial EoS, and further into representing real-fluid thermochemistry have already been well summarized in our previous studies [32], which will only be briefly discussed herein.

The $N^{th}$ order Virial EoS can be expressed as:

$$\frac{P\bar{v}}{RT} = 1 + \frac{B_2}{\bar{v}} + \cdots + \frac{B_N}{\bar{v}^{(N-1)}} \qquad (1)$$

where $B_2$, ..., $B_N$ are the second, ..., and $N^{th}$-order Virial coefficients which represent intermolecular interactions between two molecules, ..., and $N$ molecules, respectively. These Virial coefficients need to be derived for each species in the reacting mixture (i.e., the chemistry set used) and can be computed from intermolecular potentials. These intermolecular potentials are determined *ab initio* in this work whenever applicable. A complete set of the species where the Virial coefficients are determined from *ab initio* intermolecular potentials is listed in [32], which covers all the major species (e.g., fuels, oxidizers, diluents) involved in this study. The NUIGMech1.1 mechanism [37] is used as the chemistry set in this study, for which the pure-substance and cross Virial coefficients have been computed in our previous studies [33].

Based on the high-order mixture Virial EoS, real-fluid thermodynamic departure functions [32] and real-fluid chemical equilibrium [33] are rederived, based on which the governing equations for high-pressure flow reactors are further rederived. Several classical assumptions are made for the FR, as have been done in most previous modelling studies assuming plug flow [27, 28]:

i. Steady-state flow throughout the FR.
ii. No mixing in the axial direction.
iii. Constant section area.
iv. The temperature profile along the axial direction is pre-determined according to the experimental data, and therefore, there is no need to consider the conservation law of energy.
v. An ideal frictionless flow is assumed, allowing the use of the Euler equation to relate pressure and velocity.

vi. One-dimensional incompressible flow is assumed, indicating uniform properties in the section perpendicular to the flow.

Based on the assumptions above, at any cross section, a single velocity, pressure, composition, and molar volume completely characterize the flow. In an *I*-component system, the species conservation of species $i$ at the axial location $x$ is:

$$\frac{d(\dot{m}Y_i)}{dx} - \dot{w}_i M_i A = 0 \qquad i = 1, 2, \ldots, I \tag{2}$$

where, $\dot{m}$ is the mass flow rate, $Y_i$ is the mass fraction of species $i$, $\dot{w}_i$ is the net production rate of species $i$, $M_i$ is the molecular weight of species $i$, and $A$ is the reactor section area. It should be noted that the net production rates of species in Eq. 2 and the following governing equations are corrected using the real-fluid chemical equilibrium, following our previous study [33].

One can find a relation between the mass fraction $Y_i$ and mole fraction $X_i$, expressed as:

$$\frac{dY_i}{dx} = M_i \frac{X_i' \overline{M} - \overline{M}' X_i}{\overline{M}^2} \tag{3}$$

where, the superscript ' represents the total or partial derivative with respect to the axial location $x$, and $\overline{M}$ denotes the average molecular weight. $X_i'$ are also corrected by incorporating real-fluid chemical equilibrium.

Combining Eq. 2 and Eq. 3 yields:

$$X_i' - \frac{\overline{M}' X_i}{\overline{M}} - \frac{\bar{v} \dot{w}_i}{v_x} = 0 \tag{4}$$

where, $v_x$ represents axial velocity and the total derivative $\overline{M}'$ can be determined by:

$$\frac{d\overline{M}}{dx} = \sum_i M_i X_i' \tag{5}$$

Substituting Eq. 5 into Eq. 4, the expressions for the conservation of species $i$ are finally expressed as:

$$X_i' - \frac{X_i}{\overline{M}} \sum_i M_i X_i' - \frac{\bar{v} \dot{w}_i}{v_x} = 0 \qquad i = 1, 2 \ldots I \tag{6}$$

Subsequently, the mass conservation at the axial location $x$ is:

$$\frac{d(\rho v_x A)}{dx} = 0 \tag{7}$$

where, $\rho$ represents the real-fluid density, calculated by molar volume and average molecular weight, following:

$$\rho = \frac{\overline{M}}{\bar{v}} \tag{8}$$

Substituting Eq. 8 into Eq. 7 and then expanding the expression yields:

$$\frac{1}{\overline{M}} \frac{\overline{M}' \bar{v} - \bar{v}' \overline{M}}{\bar{v}} + \frac{1}{v_x} \frac{dv_x}{dx} = 0 \tag{9}$$

Substituting Eq. 5 into Eq. 9 gives the final expression for mass conservation:

$$-\frac{\bar{v}'}{\bar{v}} + \frac{v'_x}{v_x} + \frac{1}{\overline{M}}\sum_i M_i X'_i = 0 \qquad (10)$$

For one-dimension incompressible flow without considering viscosity and body force, the Euler equation is simplified as:

$$\frac{dP}{dx} + \rho v_x \frac{dv_x}{dx} = 0 \qquad (11)$$

Using Eq. 8 to replace density $\rho$ in Eq. 11 yields the final function for momentum conservation:

$$P' + \frac{\overline{M}v_x}{\bar{v}}v'_x = 0 \qquad (12)$$

As the partial derivative of molar volume $\bar{v}'$ in Eq. 9 is still unknown, the Virial EoS is utilized to derive the last governing equation. Differentiating Eq. 1 yields:

$$\frac{1}{RT^2}\left(\frac{dP}{dx}T - \frac{dT}{dx}P\right) = -\frac{\bar{v}'}{\bar{v}^2} + \frac{B'_2\bar{v}^2 - 2B_2\bar{v}\bar{v}'}{\bar{v}^4} + \ldots + \frac{B'_N\bar{v}^N - NB_N\bar{v}^{N-1}\bar{v}'}{\bar{v}^{2N}} \qquad (13)$$

Rearranging Eq. 13 gives:

$$\frac{1}{RT}\frac{dP}{dx} - \frac{P}{RT^2}\frac{dT}{dx} = -\frac{\bar{v}'}{\bar{v}^2} + \sum_{k=2}\frac{B'_k}{\bar{v}^k} - \bar{v}'\sum_{k=2}\frac{kB_k}{\bar{v}^{k+1}} \qquad (14)$$

where the total derivative of the $k^{\text{th}}$-order Virial coefficient with respect to $x$ is expressed as:

$$B'_k = \frac{dB_k}{dx} = \left(\frac{\partial B_k}{\partial T}\right)\frac{dT}{dx} + \sum_i\left[\left(\frac{\partial B_k}{\partial X_i}\right)_{T,X_{j\neq i}}\frac{dX_i}{dx}\right] \qquad (15)$$

Substituting Eq. 15 into Eq. 14, the last governing equation is finally given by:

$$\left(\frac{P}{RT^2} + \sum_{k=2}\frac{\partial B_k}{\partial T}\frac{1}{\bar{v}^k}\right)T' = \frac{1}{RT}P' + \left(\frac{1}{\bar{v}^2} + \sum_{k=2}\frac{kB_k}{\bar{v}^{k+1}}\right)\bar{v}' - \sum_{k=2}\frac{\sum_i \frac{\partial B_k}{\partial X_i}X'_i}{\bar{v}^k} \qquad (16)$$

Solving Eqs. 6, 10, 12, and 16 (which together constitute $I+3$ governing equations) gives the evolution of pressure, velocity, and mole fractions along the axial direction for real-fluid pyrolysis and oxidation in high-pressure FRs.

Additionally, to quantify the influences of real-fluid behaviors in high-pressure flow reactors, the ideal-gas modelling is also conducted using Cantera [38], indicated by the differences in the simulated species profiles between ideal-gas modelling and real-fluid modeling.

*2.2 Test conditions*

Table 1 summarizes the mixtures (i.e., hydrogen mixtures and a syngas mixture) and initial conditions (where, $\Phi$ and $D$ denote the equivalence ratio and dilution ratio (on molar basis), respectively) used in this study for real-fluid modelling of high-pressure FRs. It should be noted that Cases 1-3 are taken from previous experimental studies in high-pressure FRs (the experiment measurements are also used for comparison in the following), while Case 4 is defined with a significantly higher pressure, aiming to further elucidate the real-fluid effects in extremely high-pressure FRs. The critical pressure ($P_{cr}$) of each mixture, estimated using Cantera [38], is also included in Table 1. Furthermore, in Case 3, NO and $NO_2$ totally account for 0.015% of the mixture on the molar basis, and the specific mole fractions of each NO and $NO_2$ can be referred to [41]. As can be seen from Table 1, the conditions covered by Cases 1-4 encompass both subcritical and supercritical conditions. For all cases, the pre-determined temperature profiles along the axial direction of the FRs are extracted from [39-41] accordingly, which consist of isothermal and non-isothermal sections, with the isothermal-section temperature denoted by $T$ in Table 1. Case 4 adopts the same temperature profiles as those in Case 2.

Table 1. The conditions and experiments used in this study.

| Case No. | Mixture | $P_{cr}$ (bar) | $T$ (K) | $P$ (bar) | $\Phi$ | $D$ (%) | Ref |
|---|---|---|---|---|---|---|---|
| 1 | $H_2/O_2/N_2$ | 36.6 | 946 | 10-48 | 0.03 | 78.4 | [39] |
| 2 | $H_2/O_2/N_2$ | 34.1-49.3 | 703-899 | 50 | 0.0009-12.07 | 5.9-99.5 | [40] |
| 3 | $H_2/CO/O_2/NO/NO_2/N_2$ | 34.2 | 601-898 | 20-100 | 0.06 | 98.3 | [41] |
| 4 | $H_2/O_2/N_2$ | 34.2 | 703-899 | 500 | 1.03 | 99.5 | - |

## 3. Results and discussions

*3.1 Oxidation of hydrogen in high-pressure FRs*

3.1.1 Real-fluid effects at various pressures

Figure 1 summarizes the simulated molar concentrations of $H_2$ and $H_2O$ at the FR outlet for Case 1, along with the experimental measurements from [39]. Simulations are conducted using both the ideal modeling framework and the real-fluid modeling framework developed in this study, where the results are marked as "Ideal gas" and "Real fluid" in Fig. 1. The real-fluid effects are also quantified in Fig. 1, which is computed as the relative change in the simulated mole fraction between the ideal and real-fluid simulations. It can be seen from Fig. 1 that at the lowest pressure (i.e., P = 10 bar), the differences between the ideal and real-fluid modeling results are negligible (e.g., the relative change in simulated mole fraction is close to 0% at the lowest pressure). This indicates that the real-fluid modeling framework converges to the ideal modeling framework when real-fluid behaviors become weak (e.g., at low pressures), which validates the reliability of the developed real-fluid modeling framework. As pressure increases and real-fluid behavior becomes stronger, the simulated molar concentrations from the real-fluid modeling framework start to deviate from those from the ideal modeling framework. This is most obvious in Fig. 1a, where the simulated $H_2$ mole fraction from the real-fluid modeling framework becomes obviously lower than the ideal case, with the largest reduction observed at the highest pressure (i.e., -5.2% at P = 50 bar). This indicates that real-fluid behaviors in FRs promote the fuel oxidation reactivity, which is consistent with the trends observed in our previous studies in other fundamental reactors (e.g., shock tubes [32, 33], rapid compression machines [34], jet stirred reactors [35, 36]). The promoted fuel oxidation leads to the promoted production of oxidation intermediates and products, which is confirmed in Fig. 1b, where the $H_2O$ molar concentration becomes mostly greater after real-fluid effects are incorporated. It also can be seen from Fig. 1 that the simulation results from both frameworks agree poorly with the experiments, which is due primarily to the insufficiency of the chemistry model. It is important to recognize that such insufficiency in the chemistry model will not affect the quantification of real fluid effects, as chemical kinetics are kept identical between the ideal and real-fluid modeling. The differences observed in the simulation results between the ideal and real-fluid modeling frameworks are solely contributed by whether or not incorporating real-fluid thermodynamics and chemical equilibrium.

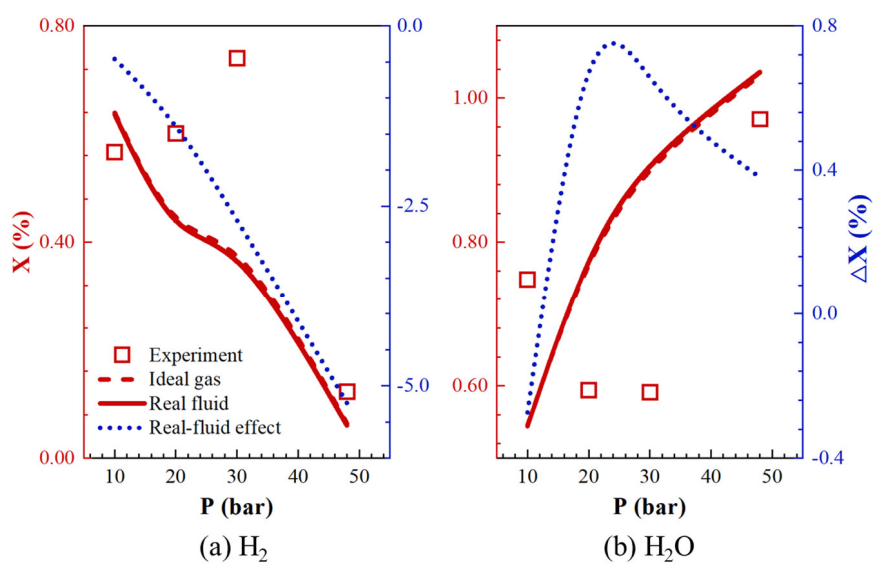

**Figure 1. Simulated mole fractions (lines) at the FR outlet for Case 1 and the respective real-fluid effects (quantified as $\Delta X = (X_{real} - X_{ideal})/X_{ideal}$), along with the experimental data (symbols) from [39]. Dashed line – simulations conducted using the ideal modeling framework; solid line – simulations conducted using the real-fluid modeling framework developed in this study.**

Figure 1 only presents the molar concentrations at the FR outlet, while those within the FR are not shown. As such, the simulated species profiles along the axial direction of the reactor from the ideal and real-fluid modeling frameworks are further shown in Fig. 2a for Case 1 at three representative pressures, along with the experimental measurements from [39]. The real-fluid effects are also quantified in a similar manner as those used in Fig. 1, which are shown in Fig. 2b. To avoid unphysical values of relative changes in Fig. 2b, the relative change profiles in Fig. 2b (as well as in the following figures) are truncated either at the start or at the end of the FR, whenever the species mole fraction reaches 1% of the respective highest value. It can be seen from Fig. 2 that both model frameworks reproduce the qualitative trends observed in the experiments. The real-fluid effect is negligible before the onset of oxidation (e.g., x<0.6 m at 10 bar), while becoming noticeable thereafter. After the onset of oxidation, the real-fluid effects promote oxidation of fuel and formation of oxidation products, with a greater impact observed at higher pressure conditions. At P = 48 bar, ignoring the real-fluid effects can underestimate the oxidation of $H_2$ by approximately 6%.

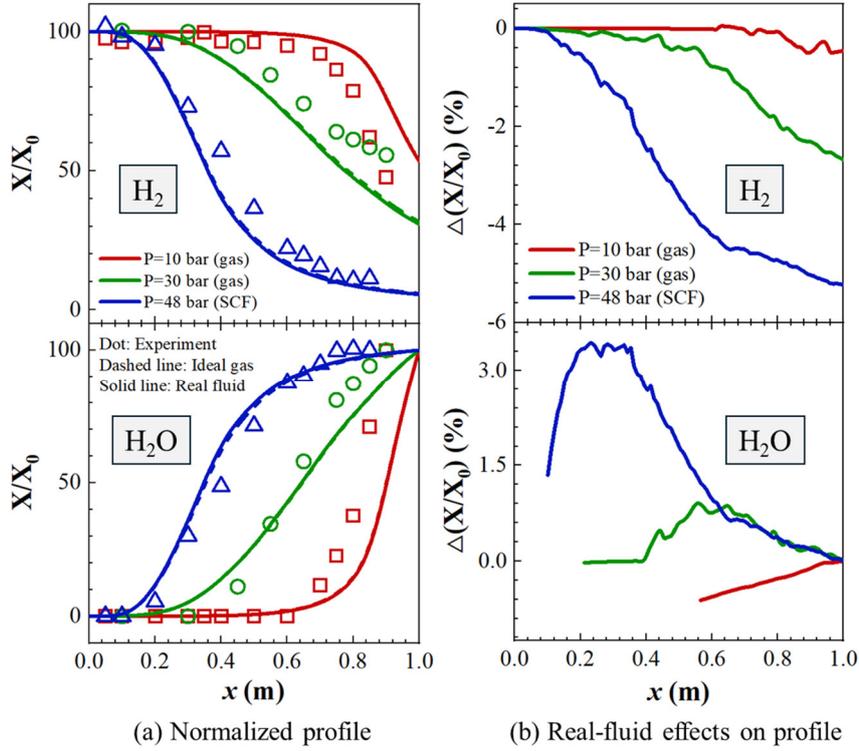

Figure 2. Normalized species profiles (lines) in the FR for Case 1 and the respective real-fluid effects (quantified as $\Delta(X/X_0) = ((X/X_0)_{real} - (X/X_0)_{ideal})/(X/X_0)_{ideal}$), along with experimental data (symbols) from [39]. $X_0$ – mole fraction of $H_2$ at the FR inlet and $H_2O$ at the FR outlet. Dashed line – simulations conducted using the ideal modeling framework; solid line – simulations conducted using the real-fluid modeling framework developed in this study.

As can be seen in Eqs. 6, 10 and 12, the axial velocity ($v_x$) is explicitly correlated with the species mole fractions. As such, the real-fluid effects on the axial velocity profiles in FRs are also quantified. The results for Case 1 are presented in Fig. 3. As can be seen from Figs. 3a and 3b that at P = 10 and 30 bar, the real-fluid effects tend to accelerate the flow, with the highest acceleration of $\Delta v_x$ = 0.74% and 0.48%, respectively, seen at around x = 0.85m. However, at P = 48 bar (Fig. 3c), an opposite trend is observed, where the flow becomes slower with real-fluid effects incorporated. Lower local speed indicates longer local residence time of the mixture in the FR, allowing more time for oxidation to take place, hence stronger real-fluid effects (as confirmed in Fig. 2). This is interesting and unique for FRs, as our previous studies in shock tubes [32, 33] and RCMs [34] indicated that the residence time for chemical kinetics becomes shorter (i.e., shorter ignition delay) with stronger real-fluid effects, which could imply a more complicated mechanism for the influences of real-fluid behaviors in FRs.

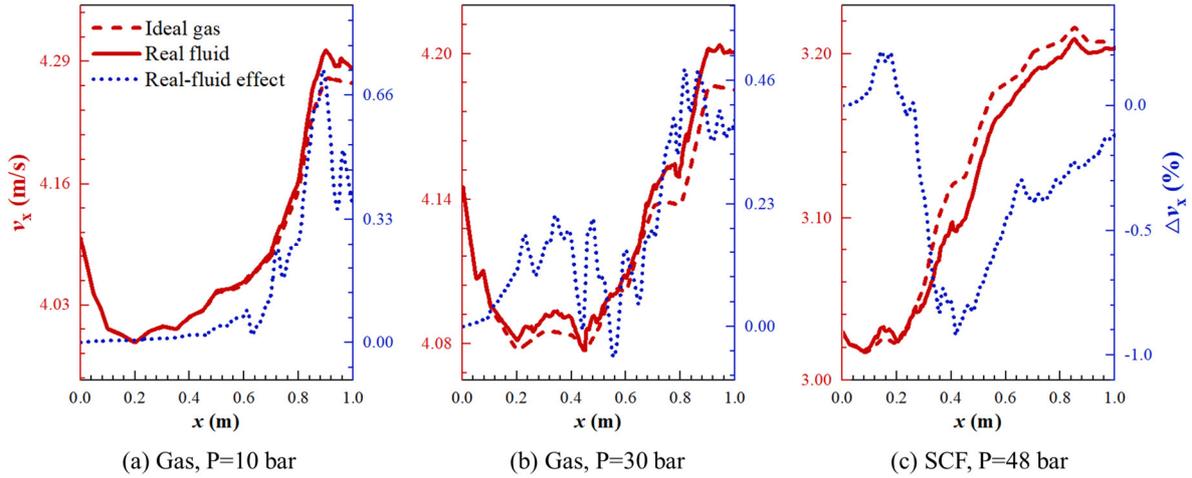

| (a) Gas, P=10 bar | (b) Gas, P=30 bar | (c) SCF, P=48 bar |

Figure 3. Simulated axial velocity profiles in the FR and the respective real-fluid effects (quantified as $\Delta v_x = (v_{x,real} - v_{x,ideal})/v_{x,ideal}$) for Case 1. Dashed line – simulations conducted using the ideal modeling framework; solid line – simulations conducted using the real-fluid modeling framework developed in this study.

3.1.2 Real-fluid effects at various temperatures and equivalence ratios

To further elucidate the real-fluid effects at various temperatures and equivalence ratios, ideal simulations and real-fluid simulations are further conducted for Case 2. The simulated species mole fractions of $H_2$, $O_2$ and $H_2O$ at the FR outlet and the quantified real-fluid effects are summarized in Fig. 4, along with the reported experiments (whenever available). It can be seen from Fig. 4 that at relatively lower temperatures where fuel (i.e., $H_2$) is barely consumed and at relatively higher temperatures where fuel oxidation is completed, the real-fluid effects on both reactants (i.e., $H_2$ and $O_2$) and products (i.e., $H_2O$) are limited and less influenced by temperature and equivalence ratio. However, under conditions where the fuel is partially oxidized through the reactor, the real-fluid effects can be greatly affected by temperature and equivalence ratio. Specifically, at temperatures from 780 K to 830 K, the promoting effects of real-fluid behavior become stronger at high temperatures at all equivalence ratios. This is most obvious for $H_2$ (Fig. 4a), where the real-fluid effects can reduce the simulated $H_2$ mole fraction by over 14% (i.e., $\Phi = 0.0009$ and T = ~840 K). The dependence of real-fluid effects on equivalence ratio is less clear, which could be complicated by the base concentration of and the balance between the reactants. Nevertheless, it can be seen from Fig. 4 that the real-fluid effects become stronger at conditions close to stoichiometry. The dependence of real-fluid effects on equivalence ratio can be better seen from the results for $H_2O$ (Fig. 4c), which has similar concentrations across different mixtures. It is clear that the real-fluid effects are significant at the conditions covered by Case 2 and become the greatest when the equivalence ratio is close to 1.0, where a change of 55.4% in the simulated mole fraction of $H_2O$ is observed at T = 775 K.

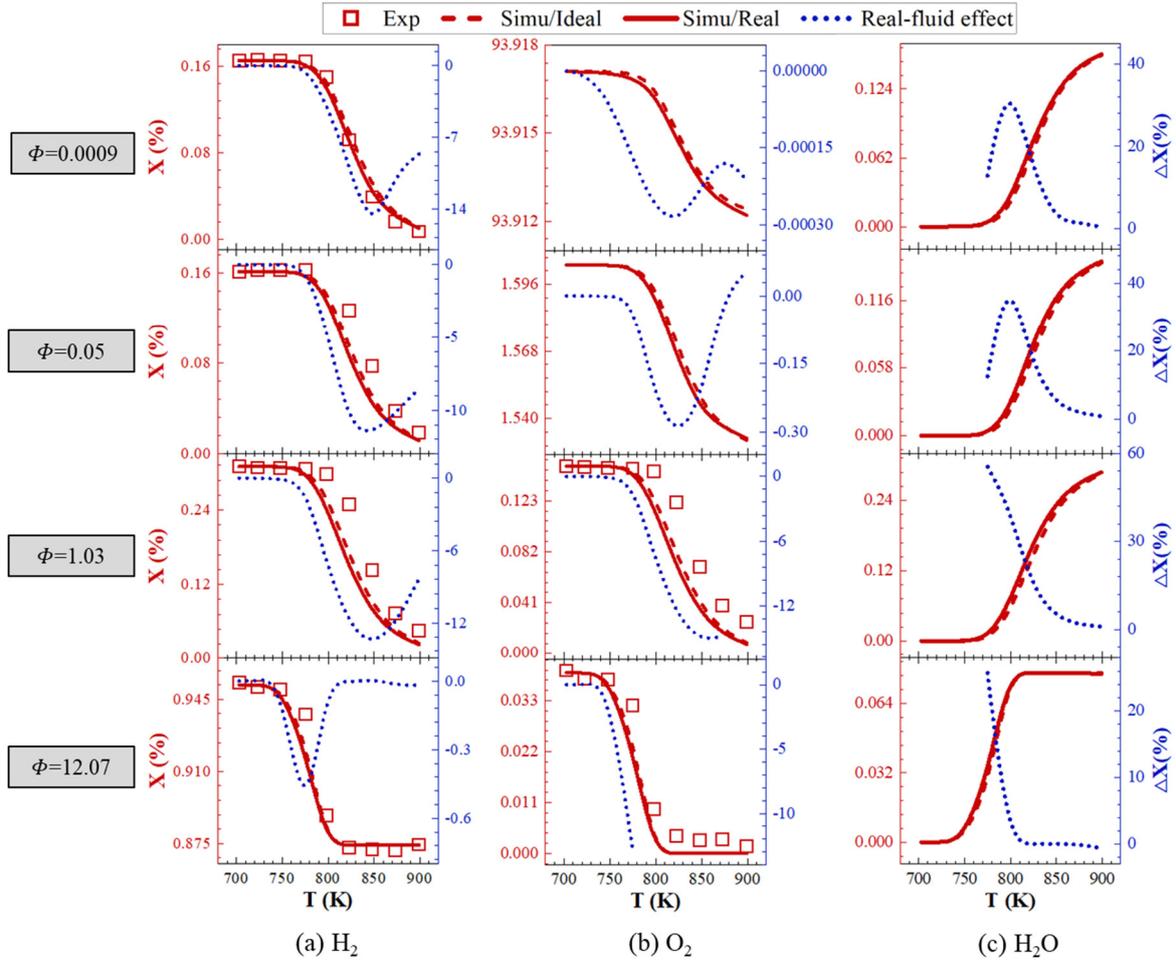

Figure 4. Simulated mole fractions (lines) at the FR outlet for Case 2 and the respective real-fluid effects (quantified as $\Delta X = (X_{real} - X_{ideal})/X_{ideal}$), along with the experimental data (symbols) from [40]. Dashed line – simulations conducted using the ideal modeling framework; solid line – simulations conducted using the real-fluid modeling framework developed in this study. T – Temperature of the isothermal section.

Figure 5 further illustrates the normalized species profiles and the respective real-fluid effects at T = 798 K and at $\Phi$ = 1.03 and 12.07 for Case 2. Note that under the conditions investigated, the isothermal section of the FR only covers 0.6 m<x<1.10 m with a temperature at 798 K, while in the non-isothermal sections (i.e., x<0.6 m and x>1.10 m), the temperature is around 460 K [40]. Thus, the mixture reactivity in the non-isothermal sections is low, as can be seen from the unchanged mole fractions in the non-isothermal sections in Fig. 5. The real-fluid effects in these sections are also negligible. However, at 0.6 m<x<1.10 m, where mixture reactivity is strong, the real-fluid effects on $H_2$ and $O_2$ consumption become strong, promoting the consumption of $H_2$ and $O_2$. At $\Phi$ = 12.07, with sufficient fuel in the mixture (e.g., a fuel-rich mixture), the real-fluid effects can affect the simulated $O_2$ mole fraction by up to 90%, whereas with comparable fuel and air (e.g., close to equivalence ratio of 1.0), the real-fluid effects impose similar relative change to the simulated $O_2$ and $H_2$ mole fractions, promoting their consumption by approximately 7%. The analyses with Fig. 5 indicate that the real-fluid effects in FRs are correlated with the mixture reactivity in the FRs, which can be affected by the temperature distribution along the axial direction in the FR. This is not obvious with Fig. 2 for Case 1, which could be due to the fact that there is no significant difference in temperature between the isothermal and non-isothermal sections in the FR for Case 1, which is within ±50 K at around 950 K [39].

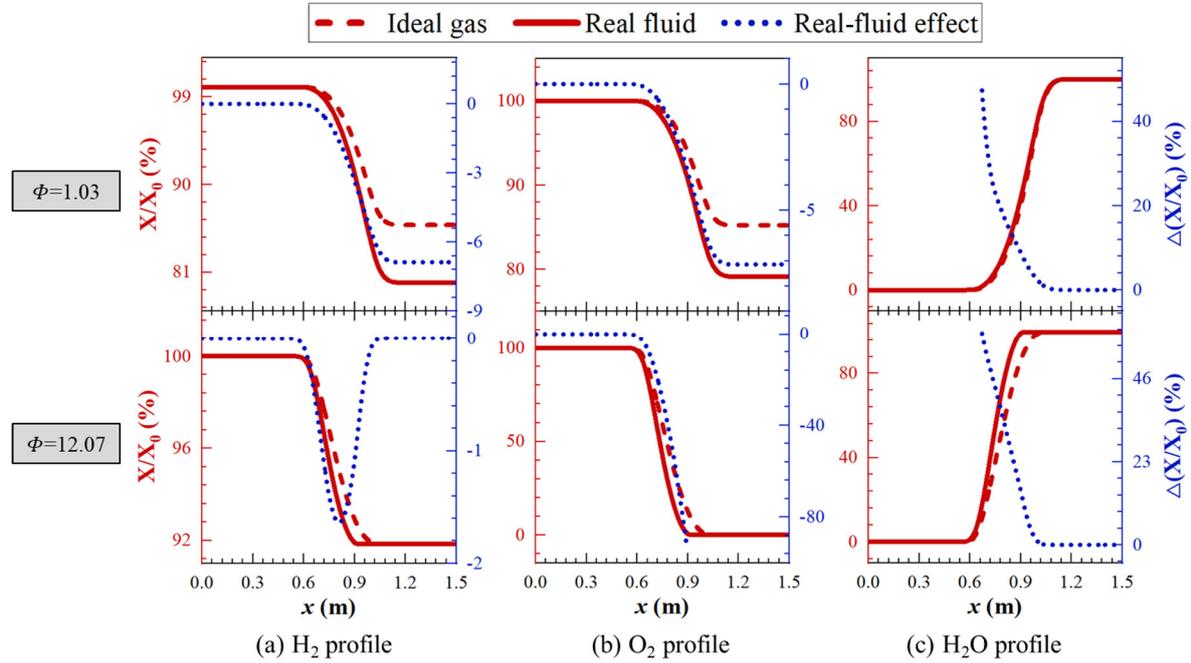

Figure 5. Normalized species profiles (lines) in the FR for Case 2 at T = 798 K and the respective real-fluid effects (quantified as $\Delta(X/X_0) = ((X/X_0)_{real} - (X/X_0)_{ideal})/(X/X_0)_{ideal})$. $X_0$ – mole fraction of $H_2$ or $O_2$ at the FR inlet and $H_2O$ at the FR outlet. Dashed line – simulations conducted using the ideal modeling framework; solid line – simulations conducted using the real-fluid modeling framework developed in this study.

*3.2 Oxidation of syngas in high-pressure FRs*

The real-fluid effects during syngas oxidation in a high-pressure FR are further investigated. Compared with the results in Section 3.1 for $H_2$ oxidation, the results in this Section can help reveal the fuel-to-fuel difference in real-fluid effects in high-pressure FRs. The simulated mole fractions at the FR outlet for Case 3 and the respective real-fluid effects are summarized in Fig. 6, along with the experimental data from [41].

First seen from Fig. 6 is the similar qualitative trends between the ideal and real-fluid simulations for the simulated species profiles with respect to the FR temperature, validating the reliability of the developed real-fluid modeling framework. It is also interesting to see that, for Case 3, the real-fluid effects inhibit the syngas oxidation reactivity through the FR, resulting in higher simulated mole fractions for the reactants (e.g., $H_2$ and CO) and lower simulated mole fractions for the major oxidation products (e.g., $CO_2$ and $H_2O$) at the FR outlet. This is opposite to the trends observed in Section 3.1, where the real-fluid effects promote $H_2$ oxidation reactivity. These opposite effects reveal the complicated real-fluid effects in high-pressure FRs, which are dependent on fuel type and mixture composition. This dependence has also been observed in other types of fundamental reactors, such as shock tubes [33] and jet-stirred reactors [35]. The reversed real-fluid effects for syngas oxidation reactivity could be attributed to several reasons: (i) the NO and $NO_2$ in the mixture of Case 3 exhibit negative compressibility factors under the influence of real-fluid behaviors [42, 43], which are opposite to those of $N_2$ and $O_2$. These contradictory real-fluid effects could weaken the promoting effects of real-fluid behaviors on oxidation reactivity that has been observed with Case 2; (ii) NO and $NO_2$ promote the generation of radicals (e.g., OH and $HO_2$ [41]). These radicals compete with the stable species in real-fluid effects and tend to inhibit mixture reactivity, as previously found in [32, 33]. The enhanced radical production with NO and $NO_2$ addition could therefore lead to an overall inhibiting real-fluid effect on oxidation reactivity. The dependence of real-fluid effects on temperature is also obvious from Fig. 6, where stronger promoting effects of real-fluid behaviors on oxidation reactivity are observed at higher pressure conditions. For instance, the oxidation of $H_2$ is inhibited by real-fluid effects by 24.2% at 820 K and 20 bar, which is further reduced by approximately 80.0% and 100.5% at 50 bar and 100 bar, respectively.

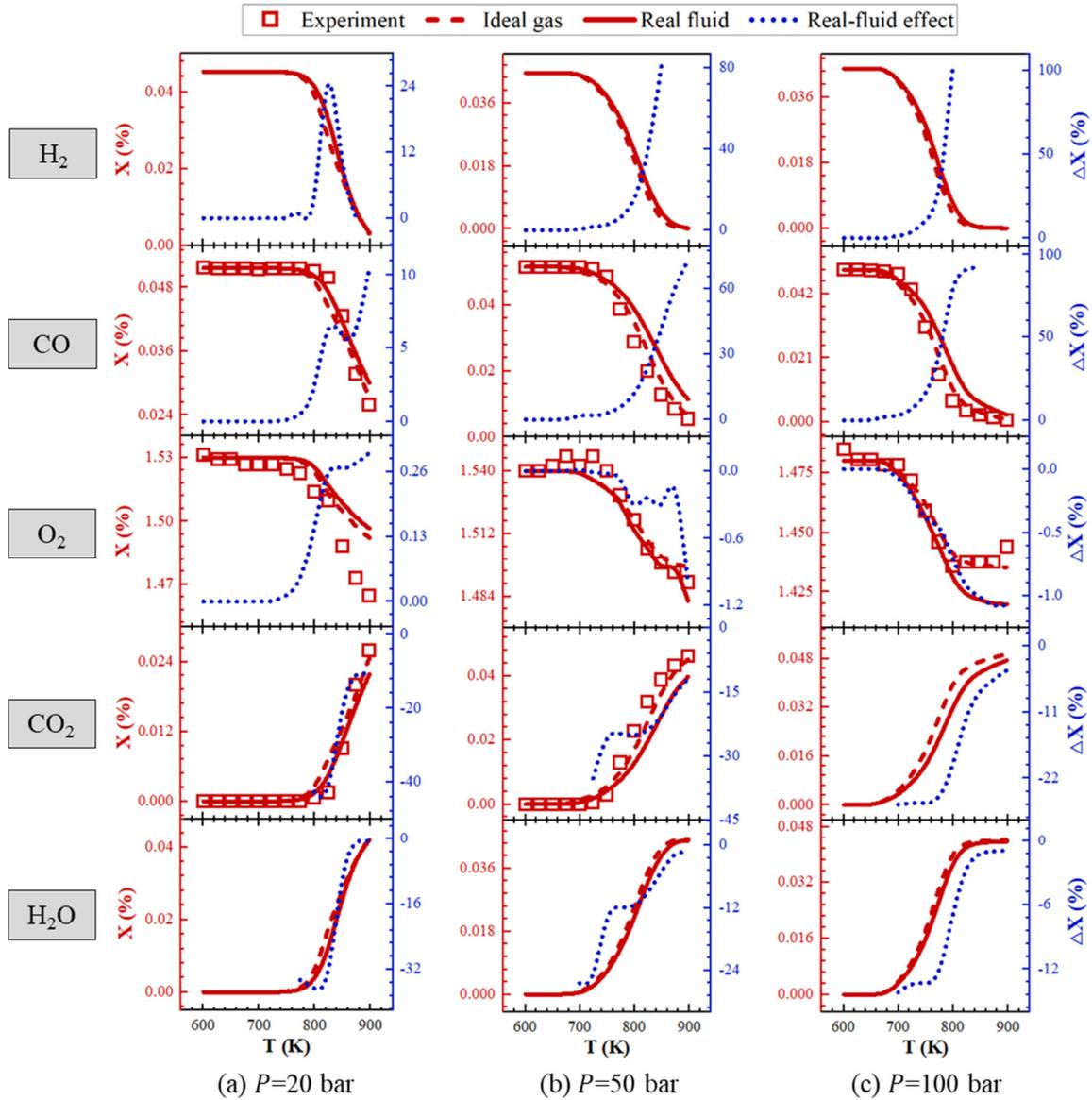

Figure 6. Simulated mole fractions (lines) at the FR outlet for Case 3 and the respective real-fluid effects (quantified as $\Delta X = (X_{real} - X_{ideal})/X_{ideal}$), along with the experimental data (symbols) from [41]. Dashed line –

To further investigate the oxidation process of syngas inside FRs, the normalized species profiles and the respective real-fluid effects at T = 775 K for Case 3 are illustrated in Fig. 7. Here, the temperature of 775 K is selected, as under the three pressures investigated, the oxidation at this temperature has begun while the reactants haven't been all consumed at the FR outlet, conducive to investigating the evolution of the real-fluid effect inside FRs. At the non-isothermal sections (i.e., x<0.33 m and x>0.79 m) where temperatures are at around 395 K, the mole fractions under all pressures are nearly unchanged, similar to the phenomenon observed in Fig. 5. At the isothermal section (0.33 m<x<0.79 m), the real-fluid effect is strengthened with increasing pressure, suppressing the consumption of $H_2$ and CO by 30% and 33% at P = 100 bar, respectively.

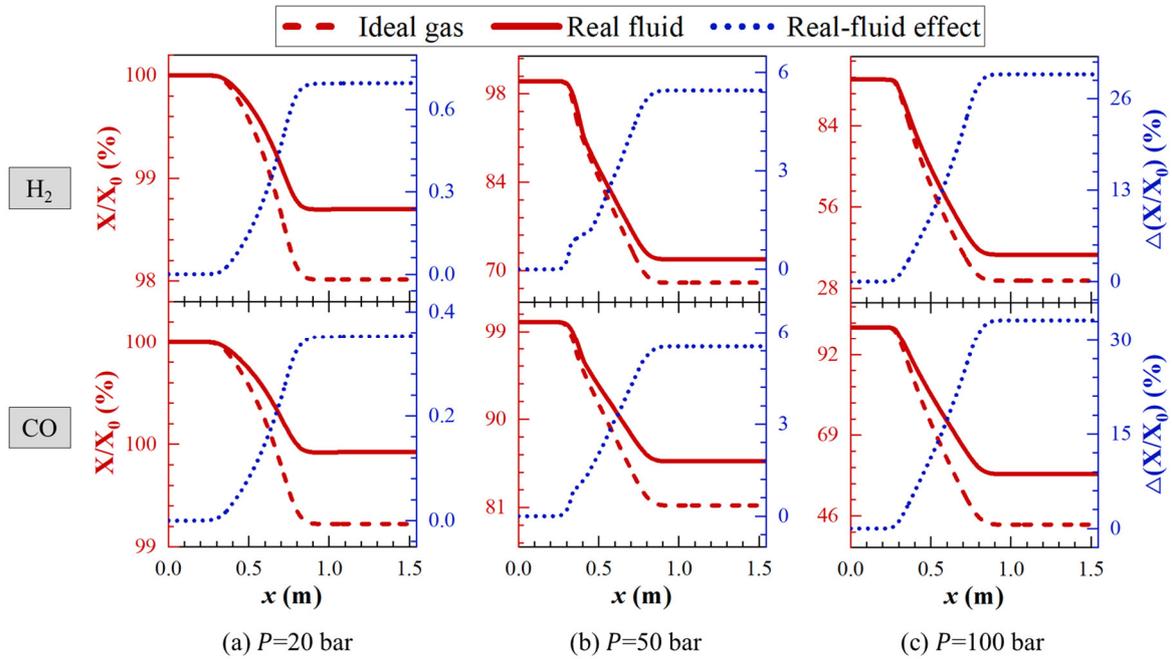

**Figure 7. Normalized species profiles (lines) in the FR for Case 3 at T = 775 K and the respective real-fluid effects (quantified as** $\Delta(X/X_0) = ((X/X_0)_{real} - (X/X_0)_{ideal})/(X/X_0)_{ideal})$. $X_0$ – **mole fraction of $H_2$ or CO at the FR inlet. Dashed line – simulations conducted using the ideal modeling framework; solid line – simulations conducted using the real-fluid modeling framework developed in this study.**

*3.3 Real-fluid effects over wider pressures*

In previous discussions concerning Cases 1-3, the real-fluid oxidation characteristics are investigated at pressures of 100 bar or below. Nevertheless, with the designing and application of ultra-high-pressure FRs, whether used for fundamental research (e.g., flow reactors [44] at above 400 bar) or industrial applications, it is necessary to investigate the real-fluid effects in high-pressure FRs at higher pressure conditions. Toward this, the simulation for Case 4 is further conducted at 500 bar, with the other initial conditions along with pre-determined temperature profiles being the same as those in Case 2 at $\Phi$ = 1.03. The simulated mole fractions at the FR outlet and the respective real-fluid effects are illustrated in Fig. 8, with the ideal-gas modelling results denoted by dashed lines while the real-fluid modelling results (referred to as "Real fluid (full)") are denoted by solid lines.

As can be seen from Fig. 8, the oxidation profiles are pushed to a lower temperature range due to enhanced reactivity at the higher-pressure condition, as compared to the results at a lower pressure in Fig. 4. Similar to those observed in Fig. 4, the real-fluid effects enhance the oxidation reactivity, leading to reduced simulated mole fractions for $H_2$ and $O_2$ and increased simulated mole fraction for $H_2O$. The real-fluid effects are strong at low temperatures when $H_2$ has not been completely consumed at the FR outlet, which can change the simulated mole fraction by over 100% for $H_2O$ and 65% for $H_2$. These changes are considerably greater than those observed in Fig. 4 at 50 bar, indicating the significant augmented real-fluid effects at higher pressure conditions.

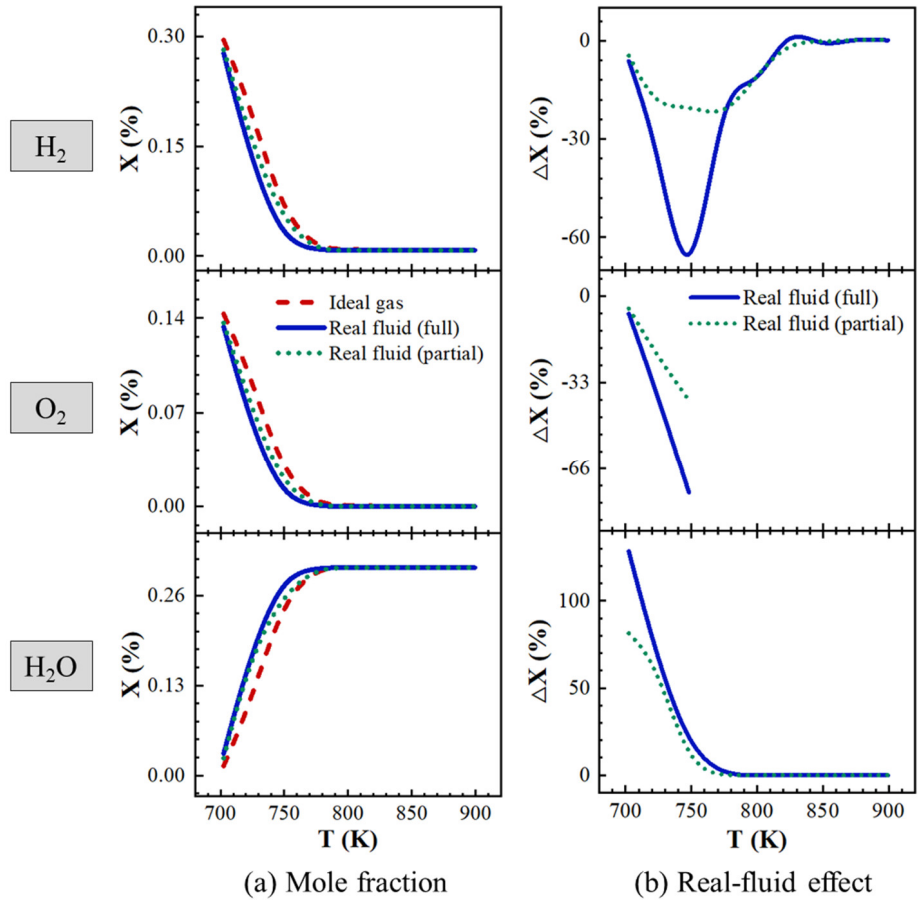

Figure 8. Simulated mole fractions (i.e., $X$) and the related quantified real-fluid effects (i.e., $\Delta X = (X_{real} - X_{ideal})/X_{ideal}$) at the FR outlet for Case 4 at $\Phi = 1.03$ from the ideal gas oxidation modelling and real-fluid oxidation modelling. full – real-fluid oxidation modelling considering all real-fluid effects. partial - real-fluid oxidation modelling ignoring real-fluid effects on chemical equilibrium. $T$ – Temperature of the isothermal section.

The normalized simulated species profiles and velocity profiles in the FR for Case 4 at T = 775 K, along with the quantified real-fluid effects on them, are further shown in Fig. 9. As can be seen from Fig. 8, at T = 775 K, the oxidation of $H_2$ has almost been completed. Although the real-fluid effects do not affect the mole fractions at the FR outlet at this temperature, the progression of oxidation within the reactor might still be affected by the real-fluid effects. This is confirmed in Figs. 9(a) and 9(b), where the species profiles in the isothermal section of the FR (i.e., 0.6 m<x<1.1 m) display different trends between the ideal modeling results and the real-fluid modeling results. It is clear from Fig. 9(a) that real-fluid effects accelerate the oxidation process in the FR, as $H_2$ is completely consumed at a shorter distance (i.e., x = 0.84 m) from the FR inlet in the real-fluid modeling case, as compared to that (i.e., x = ~1.1 m) in the ideal case. The quantitative change in simulated species mole fractions after incorporating the real-fluid effects at a specific location in the FR is significant, reaching $\Delta(X/X_0)$ = -64.9% and -85.0% for $H_2$ and $O_2$, respectively.

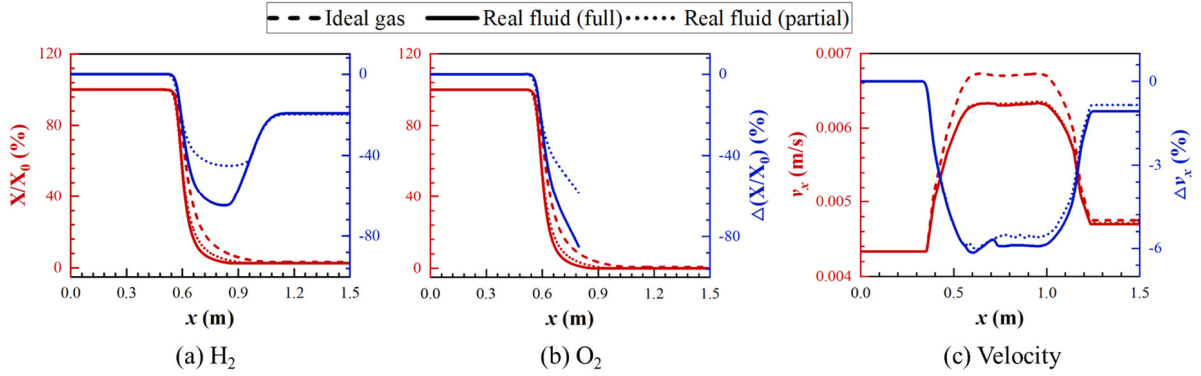

(a) $H_2$  (b) $O_2$  (c) Velocity

**Figure 9.** Normalized species profiles (i.e., $X/X_0$) and velocity profiles (i.e., $v_x$), and the related quantified real-fluid effects (i.e., $\Delta(X/X_0) = ((X/X_0)_{real} - (X/X_0)_{ideal})/(X/X_0)_{ideal}$, $\Delta v_x = (v_{x,real} - v_{x,ideal})/v_{x,ideal}$) at $T$ = 775 K and $\Phi$ = 1.03 simulated through the ideal-gas oxidation modelling and real-fluid oxidation modelling for Case 4. $X_0$ – mole fraction of $H_2$ or $O_2$ at the FR inlet. full – real-fluid oxidation modelling considering all real-fluid effects. partial – real-fluid oxidation modelling ignoring real-fluid effects on chemical equilibrium.

*3.4 Decoupling between real-fluid chemical equilibrium and real-fluid thermodynamics*

The real-fluid modeling results presented in Figs. 1-7 incorporate both the real-fluid thermodynamics and real-fluid equilibrium, where the real-fluid equilibrium is calculated based on real-fluid partial fugacity coefficients that have been rederived based on the high-order Virial EoS [33]. To investigate the individual contribution of real-fluid thermodynamics and real-fluid equilibrium to the overall real-fluid effects observed in FRs, real-fluid simulations where only the real-fluid thermodynamics are considered are also conducted for Case 4. This is achieved by reverting the partial fugacity coefficients to those determined based on the ideal-gas assumption. The results are presented in Figs. 8 and 9, with the results labelled "Real fluid (partial)". It can be seen that without real-fluid equilibrium, the overall real-fluid effects are greatly suppressed. This is most obvious in Fig. 8, where the maximum quantified real-fluid effects are reduced by 64.6%, 48.0%, and 36.7% for $H_2$, $O_2$, and $H_2O$, respectively. Similar trends are also observed in Fig. 9 within the FR where real-fluid equilibrium accounts for approximately half of the total real-fluid effects. These results demonstrate the significant contribution from real-fluid equilibrium, which should be considered when modeling high-pressure flow reactor experiments.

*3.5 Implications of real-fluid behaviors on the validation of chemistry models against measurements in high-pressure FRs*

In previous fundamental combustion studies using high-pressure flow reactors [9-24, 39-41], measured species profiles were typically used for comparison against simulation results, thereby validating and fine-tuning the chemistry models used for FR simulations. When disagreements were observed between the experimental measurements and simulation results, adjustments were always made to the chemistry model such that the simulation results can better match the experiments, while the potential error in the modeling framework for FRs has been completely overlooked. In all past FR studies in the fundamental combustion community, the modeling of FRs has been conducted via Chemkin [29], Cantera [38], or other in-house codes, which mainly assume ideal gas behavior in the FRs. This is reasonable at atmospheric to low-pressure conditions, but might become problematic at high-pressure conditions where real-fluid behaviors become strong and the ideal gas assumption fails. This is directly confirmed in this study with the results in Figs. 1-8. For the experimental conditions investigated in this study (i.e., Case 1-3), ignoring the real-fluid effects in high-pressure FRs leads to maximum deviations of -16.6%, -15.2%, and 55.4% in simulated mole fractions for $H_2$, $O_2$, and $H_2O$, respectively, at the FR outlet for $H_2$ oxidation at 50 bar (c.f., Fig. 4), while for syngas oxidation in high-pressure FRs, the maximum deviations in simulated mole fractions for $H_2$, $CO$, $CO_2$, $H_2O$ at the FR outlet are 24.2%, 10.5%, -46.5%, and -37.8%, respectively, at 20 bar and 100.5%, 92.4%, -26.4% and -14.2%, respectively, at 100 bar (c.f., Fig. 6). Ignoring the real-fluid effects also shifts the species profiles within the FRs whenever oxidation takes place, regardless of whether the oxidation has been completed at the FR outlet or not. Such shifts can lead to up to 90% change in relative simulated mole fractions within the FR (c.f., Fig. 5). It is important to recognize that these deviations and shifts are a form of error caused by the inadequacy of the ideal modeling framework in capturing the real-fluid behaviors, rather than a source of modeling uncertainty. These errors observed at 20-100 bar for hydrogen and syngas oxidation are well beyond the typical level of measurement uncertainties in FRs, which have been conservatively estimated to be within 6% in measured mole fractions [39-41, 44, 45]. Propagating these errors in simulated species profiles can lead to misunderstanding or misinterpretation of the fundamental oxidation chemistry in high-pressure FRs and pose significant errors in the developed chemistry models. With such errors, the developed chemistry models are certainly ill-conditioned at high-pressure conditions, and their reliability for high-pressure fundamental combustion modeling becomes questionable. On the other hand, the non-ideal physics involved in high-pressure combustion processes should be replicated by a modeling framework by all means (provided within feasible computational cost), even if it is not used for developing or validating chemistry models. Unfortunately, this has not been adequately achieved in the past with regard to replicating real-fluid behaviors and their influences in high-pressure combustion, as has been confirmed in this study for high-pressure FRs and in our previous studies for other fundamental reactors, including RCMs [34], Shock tubes [32, 33], JSRs [35, 36]. It is also interesting and important to see that these real-fluid behaviors already become pronounced even at subcritical conditions (e.g., at 20 bar for Case 3 in this study, which is below the critical pressure of the mixture (c.f., Table 1)). This further highlights the necessity to sufficiently incorporate real-fluid behaviors for modeling high-pressure flow reactors (not necessarily supercritical flow reactors), which has now become possible with the real-fluid modeling framework developed in this study.

**Conclusions**

Flow reactors have been increasingly used in the fundamental combustion community to probe into fuel pyrolysis and oxidation chemistry at high-pressure conditions. This type of reactor has also been fundamental to the development and validation of chemistry models via chemical kinetic modeling. However, previous chemical kinetic modeling for pyrolysis and oxidation in high-pressure flow reactors has almost been completely conducted based on ideal gas assumption, although the ideal gas assumption might fail at high-pressure conditions when real-fluid behaviors become pronounced. Elucidating whether this is the case is urgent given the high demand for high-pressure combustion and propulsion technologies. Unfortunately, this has been completely overlooked in the past. Through the present study, we have provided the first-of-its-kind framework, results and knowledge that reveal the real-fluid behaviors in high-pressure flow reactors and their influences. Specifically,

1. This study successfully establishes a completely new real-fluid modeling framework for high-pressure flow reactors that fully incorporates the real-fluid behaviors in high-pressure flow reactors. This is achieved by physically representing the real-fluids at the molecule level via incorporating *ab initio* intermolecular potential into high-order Virial EoS and further coupling this with real-fluid thermochemistry, real-fluid chemical equilibrium, and real-fluid conservation laws of species, mass, and momentum.
2. With the developed real-fluid modelling framework, the real-fluid effects on the oxidation are quantified and analyzed via case studies on the oxidation of hydrogen and syngas, covering conditions P = 10-500 bar, T = 601-946 K, $\Phi$ = 0.0009-12.07, and D = 5.9%-99.5%. The real-fluid effects are found to be affected by mixture reactivity, depending on the temperature, pressure, and equivalence ratio conditions. For hydrogen oxidation, the real-fluid effects are found to promote reactant consumption (e.g., by up to 14.0% and 65.0% for $H_2$ at P = 50 bar and 500 bar, respectively) and products generation (e.g., by up to 55.4% and over 100% for $H_2O$ at P = 50 bar and 500 bar, respectively), while for the oxidation of syngas with NOx addition, reversed real-fluid effects that inhibit reactant consumption (e.g., by up to 24.2%, 80.0% and 100.5% for $H_2$ at 20 bar, 50 bar, and 100 bar, respectively) are observed.
3. The quantified real-fluid effects reveal that the errors introduced into the modeling of high-pressure flow reactors without incorporating real-fluid behaviors can be significant (e.g., -16.6–55.4% for hydrogen oxidation and -46.5%–100.5% for syngas oxidation) and are substantially greater than typical levels of measurement uncertainties in flow reactors, which might lead to misinterpretation of the fundamental oxidation chemistry in high-pressure flow reactors and posing significant errors in the developed chemistry models.

4. Decoupling of real-fluid effects indicates that real-fluid equilibrium accounts for approximately 50% of the total real-fluid effects in high-pressure flow reactors, highlighting the significance of incorporating real-fluid equilibrium, in addition to real-fluid thermodynamics, when modelling high-pressure flow reactors.

**Acknowledgments**

The work described in this paper is supported by the Research Grants Council of the Hong Kong Special Administrative Region, China under PolyU P0046985 for ECS project funded in 2023/24 Exercise, the Otto Poon Charitable Foundation under P0050998, the National Natural Science Foundation of China under 52406158, the Chief Executive's Policy Unit of HKSAR under the Public Policy Research Funding Scheme (2024.A6.252.24B), and the Natural Science Foundation of Guangdong Province under 2023A1515010976 and 2024A1515011486.

**Declaration of Competing Interests**

The authors declare no competing interests.